\documentclass[fleqn,12pt,twoside]{article}
\usepackage{espcrc1}
\usepackage{epsfig}

\title{Hadronic and Quark-Gluon Excitations of Dense and Hot
Matter\thanks{Invited talk presented by W. Weise at the 3rd
International Conference on Perspectives in Hadron Physics, ICTP,
Trieste, 7-11 May 2001; Work supported in part by BMBF, GSI and DFG.}}

\author{T. Renk\address{Physik Department, 
Technische Universit\"{a}t M\"{u}nchen,
D-85747 Garching, Germany}, 
R.A. Schneider$^a$  
and W. Weise$^a$\address{ECT*, Villa
Tambosi, I-38050 Villazzano (Trento), Italy}}

\begin{document}
% typeset front matter
\maketitle

\begin{abstract}
We summarize recent developments in our understanding of low-mass
quark-antiquark excitations in hadronic matter under various different
conditions. This includes the thermodynamics of the chiral condensate, pions as
Goldstone bosons in normal nuclear matter, and excursions into extreme
territory of the QCD phase diagram: lepton pair production from a fireball
expanding through the transition boundary between the quark-gluon and hadron phases
of QCD.
\end{abstract}

\section{Phases of QCD}
Exploring the QCD phase diagram is undoubtedly one of the great challenges in
the physics of strong interactions. At temperatures exceeding $T_C \simeq
\Lambda_{QCD} \sim$ 0.2 GeV, one expects a plasma of quarks and gluons released
from their confining forces. At $T < T_C$ and at moderate baryon densities,
the relevant QCD degrees of freedom are color-singlet hadrons. Chiral symmetry
is spontaneously broken. The vacuum is a condensate of scalar quark-antiquark
pairs. Pions act as Goldstone bosons. Their decay constant, $f_{\pi} =$ 92.4 MeV, determines the
chiral scale $4 \pi f_{\pi} \sim 1$ GeV which governs the low mass
hadron spectrum. The lightest vector mesons $(\rho, \omega)$ can be seen as the
lowest $q \bar{q}$ "dipole" excitations of the QCD vacuum. Current algebra
combined with QCD sum rules \cite{GOL} connects their masses directly with the
chiral scale, $\sqrt{2} m_V = 4 \pi f_{\pi}$, to leading order (and in the large
$N_c$ limit).

Evidently, investigating the changes of the spectral distributions of
pseudoscalar and vector (as well as axial vector) excitations of the QCD vacuum
with changing temperatures and baryon densities, from moderate to extreme, is a
key to understanding QCD thermodynamics, its phases and symmetry breaking
patterns.
\section{Chiral thermodynamics: selected topics}
A central point in the discussion of the QCD phase diagram is the transition from
the Nambu-Goldstone realization of chiral symmetry (with non-zero condensate
$\langle \bar{q} q \rangle$) to the "restored" Wigner-Weyl realization in which
the chiral condensate vanishes. Lattice QCD indicates that chiral restoration
is probably linked to the transition between composite hadrons and deconfined
quarks and gluons.

Spontaneous chiral symmetry breaking (together with explicit breaking by the
small $u$- and $d$-quark masses, $m_{u,d} < 10$ MeV) implies the PCAC or
Gell-Mann, Oakes, Renner (GOR) relation, $m^2_{\pi} f^2_{\pi} = - m_q \langle
\bar{q} q \rangle$, to leading order in the quark mass $m_q = (m_u + m_d)
/2$. The GOR relation continues to hold \cite{THOR} in matter at finite
temperature $T< T_C$ and baryon density $\rho$, when reduced to a statement
about the {\it time} component of the axial current. For this component we
denote the corresponding in-medium pion decay constant by $f^*_{\pi} (\rho,
T)$. One finds \cite{THOR}
\begin{equation}
f^{*2}_{\pi} (\rho, T) m^{*2}_{\pi} = - m_q \langle \bar{q} q \rangle_{\rho, T}
+...,
\end{equation}
where $\langle \bar{q} q \rangle_{\rho, T}$ now stands for the T- and
$\rho$-dependent chiral condensate. The "melting" of this condensate by heat or
compression therefore translates primarily into an in-medium change of the
(timelike) decay constant of the pion, given that its in-medium mass
$m^*_{\pi}$ is not much different from $m_{\pi}$ in vacuum, because of its
Goldstone boson nature.

The chiral scale, $4 \pi f^*_{\pi} (\rho, T)$, is expected to decrease when
thermodynamic conditions change toward chiral restoration. This scale defines a
gap in the low-energy hadron spectrum. A decreasing chiral gap should imply
characteristic observable changes in the meson mass spectrum.

Fig.~\ref{hadron_spectra} 
shows typical examples of calculated vector meson spectral distributions
in nuclear matter at $T = 0$ and, in contrast, their evolution with temperature
at zero baryon density. Such spectra are used in calculations of lepton pair
production rates from ultrarelativistic heavy-ion collisions as described in
later sections. The main part of this presentation focuses on the quest for
signals of the QCD equation of state through dilepton radiation from hot
matter. Before exploring such extreme territory, we summarize two topics of
current interest in chiral thermodynamics at more moderate conditions.
\subsection{Thermodynamics of the chiral condensate}
Suppose we start from a chiral effective Lagrangian, appropriate for the
hadronic phase of QCD, with Goldstone bosons (pions) coupled to baryons
(nucleons). Let ${\cal Z}$ be the partition function derived from this theory,
and $\mu$ the baryon chemical potential. The pressure is $p (\mu, T) = (T/V) ln {\cal Z}$, where $V$ is the
volume. Given this equation of state, a variant of the Hellmann-Feynman theorem
(with the quark mass treated formally as an adiabatic parameter) in combination
with the GOR relation leads to the following density and temperature dependent
chiral condensate \cite{WW01}:
\begin{equation}
\frac{\langle \bar{q} q \rangle_{\rho, T}}{\langle \bar{q} q \rangle_0} = 1 +
\frac{d p (\mu, T)}{f^2_{\pi}dm^2_{\pi}} = 1 + \frac{1}{f^2_{\pi}} \left[
\frac{\partial p (\mu, T)}{\partial m^2_{\pi}} -
\frac{\sigma_N}{m^2_{\pi}} \rho_S (\mu, T)\right],
\end{equation}
with the baryon and scalar densities, $\rho = \partial p / \partial \mu$ and
$\rho_S = - \partial p/ \partial M$, and the nucleon sigma term $\sigma_N = m_q
\partial M / \partial m_q \simeq 0.05$ GeV, where $M$ is the nucleon mass.

At $T = 0$ we have $m_q [\langle \bar{q} q \rangle_{\rho} - \langle \bar{q} q
\rangle_0 ] = \sigma_N \rho_S$. At low baryon density where $\rho_S \simeq
\rho$, the magnitude of the condensate decreases linearly with density. This
behaviour persists up to about the density of normal nuclear
matter. Detailed calculations using self-consistent thermal effective field
theory at the two-loop level \cite{WW01,KSW01} show that effects non-linear in
$\rho$ take over at higher density such that the rapid decrease of $|\langle
\bar{q} q \rangle_{\rho} |$ is slowed down.

The temperature dependence of the condensate at $\rho = 0$ is determined by
thermal pion fluctuations. The result of our recent calculations
\cite{WW01,KSW01} is close to that found in chiral perturbation theory
\cite{GER} and quite similar to the result of lattice QCD \cite{BOY}, with a
critical temperature $T_C \simeq 180$ MeV (for two flavours). The tendency towards "chiral restoration" indicated by the dropping condensate
should have observable consequences, not only at extreme densities and
temperatures, but also already in normal nuclear systems.
\subsection{Pionic $s$-waves in the nuclear medium}
Let us have a brief look at Goldstone bosons in the medium. In the exact chiral
limit with $m_q = 0$, pions are massless, and this feature persists at all
temperatures and densities as long as one stays within the Nambu-Goldstone
phase of spontaneously broken chiral symmetry. Explicit symmetry breaking
causes characteristic deviations from this pattern. They become particularly
interesting in highly asymmetric nuclear matter, with large excess of neutrons
over protons.

There has been a recent revival of interest in $s$-wave pion-nucleus
interactions, for several reasons. First, the observation of deeply bound
pionic atom states in $Pb$ isotopes \cite{GIL} has sharpened the quantitative
constraints on the $s$-wave optical potential. Secondly, there is renewed
interest in the theoretical foundations of this optical potential from the
point of view of chiral dynamics \cite{WW01,WBW97,KW01}.

To leading chiral order, at $T = 0$ and in the low density limit, the
self-energy for a $\pi^+$ or $\pi^-$ at low energy $\omega$ and momentum
$\vec{q} = 0$ is simply given by the Weinberg-Tomozawa theorem:
\begin{equation}
\Pi^{\pm} (\omega, \vec{q} = 0) = \pm \frac{\omega}{2f^2_{\pi}} (\rho_p -
\rho_n) +...,
\end{equation}
so that the primary medium effect is a splitting of the $\pi^+$ and $\pi^-$
masses, $\Delta m (\pi^{\pm}) = \pm (\rho_p - \rho_n) /4 f^2_{\pi}$, in
asymmetric nuclear matter. Systematic calculations of the $\pi^+$ and $\pi^-$
mass shifts up to two-loop order using in-medium chiral perturbation theory
\cite{KW01} predict $\Delta m (\pi^-) \simeq 14$ MeV and $\Delta m (\pi^+)
\simeq -1$ MeV at nuclear matter density  and at a neutron-to-proton ratio $N/Z
= 1.5$ characteristic of the $Pb$ region. The repulsive shift found for the
$\pi^-$ mass is however only half of what is required  by the data on deeply
bound ($1s$ and $2p$) states in pionic $Pb$ \cite{GIL}. Several options to
repair this discrepancy exist. We believe \cite{WW01} that an appealing
one is the reduction of the pion decay constant in medium, following
eq. (1). The replacement $f_{\pi} \to f^*_{\pi}$ in the pion self-energy comes
naturally because in-medium chiral perturbation theory must now be performed
relative to a modified vacuum with shifted chiral condensate. Systematic investigations of deeply bound pionic atoms with
various isotopic chains of heavy nuclei will provide further insights into
these challenging questions.
\section{Dilepton production rates and thermal QCD}
Lattice simulations with three light quark flavours indicate that QCD undergoes a phase
transition at a temperature $T_C \sim$ 150 MeV \cite{FK00}.
It is hoped that it is possible to create the quark-gluon plasma (QGP) phase
above $T_C$ in ultrarelativistic heavy-ion collisions at CERN
and RHIC. Dileptons ($e^+e^-$ and $\mu^+ \mu^-$ pairs) are interesting probes in this context since
they do not interact strongly but escape unthermalized from the hot and dense region formed in
such collisions, the fireball. As the fireball expands, it cools off and hadronization sets in at $T_C$. In 
the hadronic phase, the main dilepton sources are pion and
kaon annihilation processes which are enhanced due to the formation of the 
light vector mesons $\rho,
\omega$ and $\phi$.

The differential dilepton emission rate from a hot domain in thermal equilibrium is given by
\begin{equation}
\label{E-Rate} \frac{dN}{d^4xd^4q} = \frac{\alpha^2}{\pi^3q^2}\frac{Im \bar{\Pi}(q,T)}{e^{q^0/T}
-1}
\end{equation}
where $\bar{\Pi}$ denotes the spin-averaged time-like self energy of a photon in the heat bath, $q$ is
the photon four-momentum, $\alpha = \frac{e^2}{4\pi}$, and the lepton masses 
have been neglected. Later we compare our results with the CERES/NA45 data
taken in Pb-Au collisions at 160 AGeV (corresponding to
a c.m. energy of $\sim 20$ AGeV). This is done by integrating
eq.(\ref{E-Rate}) over the space-time history of the collision and, 
taking into account the experimental
detector acceptance,  over the
transverse momentum $p_T$ and average over the rapidity $\eta$.
\subsection{Quasiparticle description of the quark-gluon phase}
Thermal QCD perturbation theory is presumably not applicable to evaluate
Im$\bar{\Pi}$, even at high temperatures. A way to proceed is to use input
from finite temperature lattice
simulations. We have shown recently \cite{RW01} that it is possible to describe the equation of state (EoS) of such
systems to a very good approximation in terms of a non-interacting gas of
quasiparticles with thermally generated masses, incorporating confinement
phenomenologically by a temperature-dependent effective number of active degrees of
freedom.

From asymptotic freedom, we expect that at extremely high temperatures the plasma
consists of quasifree quarks and gluons. Perturbative calculations
find spectral functions of the form
$\delta(\omega^2 - k^2 - m^2(T))$ with $m(T) \sim g_s T$. 
As long as the spectral functions of
the thermal excitations at lower temperatures still resemble qualitatively this
asymptotic form, a quasiparticle description is expected to be
applicable. The thermal excitations can then be described by a 
dispersion relation $\omega^2(k) = k^2 + m^2 \, (T)$.

We assume that the thermal quark and gluon quasiparticle masses still behave as
$\tilde{g} (T) T$, with an effective coupling strength parametrized as
\begin{equation}
\tilde{g} (T) \simeq g_0 \left( 1- \frac{T_C}{T} \right)^{\gamma} \, .
\end{equation}
This form is guided by lattice results which indicate that the phase transition
is weakly first order or second order, suggesting approximately $m \sim (T - T_C)^{\gamma}$
with some "pseudocritical" exponent $\gamma$.

The second important element in this model is a temperature-dependent
confinement factor which reduces the number of thermally active degrees of
freedom as $T$ approaches $T_C$. In practice, the quasiparticle parameters are determined by reproducing lattice results
for the entropy density. The pressure and the energy density then follow
accordingly and are in excellent agreement with lattice data whenever a
comparison can be made \cite{RW01}.
%\begin{figure}[h]
%\begin{center}
%\epsfig{file=2_1_flav.eps,width=7.5cm, angle = -90}
%\end{center}
%\caption{Pressure, energy and entropy density for two light quark flavours ($m%_{u,d} = 0$) and a heavier strange quark
%($m_s \simeq 170 $ MeV) in our quasiparticle model \cite{RW01}. The arrow indic%ates the ideal gas limit of massless
%three-flavour QCD.}
%\label{2_1_flav}
%\end{figure}

In the quark-gluon phase, the basic mechanism that produces $e^+ e^-$ pairs is $q \bar{q}$ annihilation. Assume now that these quarks and antiquarks are
thermal quasiparticles, and that these quasiparticles couple to photons in
the same way as bare quarks. We can then use the standard one-loop result for $Im
 \bar{\Pi}$, with bare quark masses replaced by the thermal masses, $m_q
(T)$. The essential QCD dynamics is supposed to be incorporated in these
quasiparticle masses as they are in accordance with the lattice QCD equation of
state.

\subsection{The hadronic phase}
Below $T_C$, confinement sets in and the effective degrees of freedom
change to colour singlet, bound $q\bar{q}$ or $qqq$  states. The photon couples now to the lowest-lying
'dipole' excitations of the vacuum, the hadronic $J^P = 1^-$ states: the
$\rho$, $\omega$ and $\phi$ mesons and multi-pion states carrying the  same
quantum numbers. The electromagnetic current-current correlation function can
be connected to the currents generated by these mesons using an effective
Lagrangian which approximates the $SU(3)$ flavour sector of QCD at low
energies. We use the {\em improved Vector
Meson  Dominance} model combined with chiral dynamics of pions and kaons as
described in \cite{KKW1}. Within  this model, the following relation between
the imaginary part of the irreducible photon self-energy  $\mbox{Im}
\bar{\Pi}$ and the vector meson self-energies $\Pi_V(q)$ in vacuum is derived:
\begin{equation}
\mbox{Im} \bar{\Pi}(q) = \sum \limits_V \frac{\mbox{Im}
\Pi_V(q)}{g_V^2} \ |F_V(q)|^2. \label{ImBarPi}
\end{equation}
Here $g_V$ is the $\gamma V$ coupling constant and $F_V$ is the form factor for the coupling of
the vector meson to the multipion and $K \bar{K}$ continuum.

\begin{figure}[h]
\begin{center}
\epsfig{file=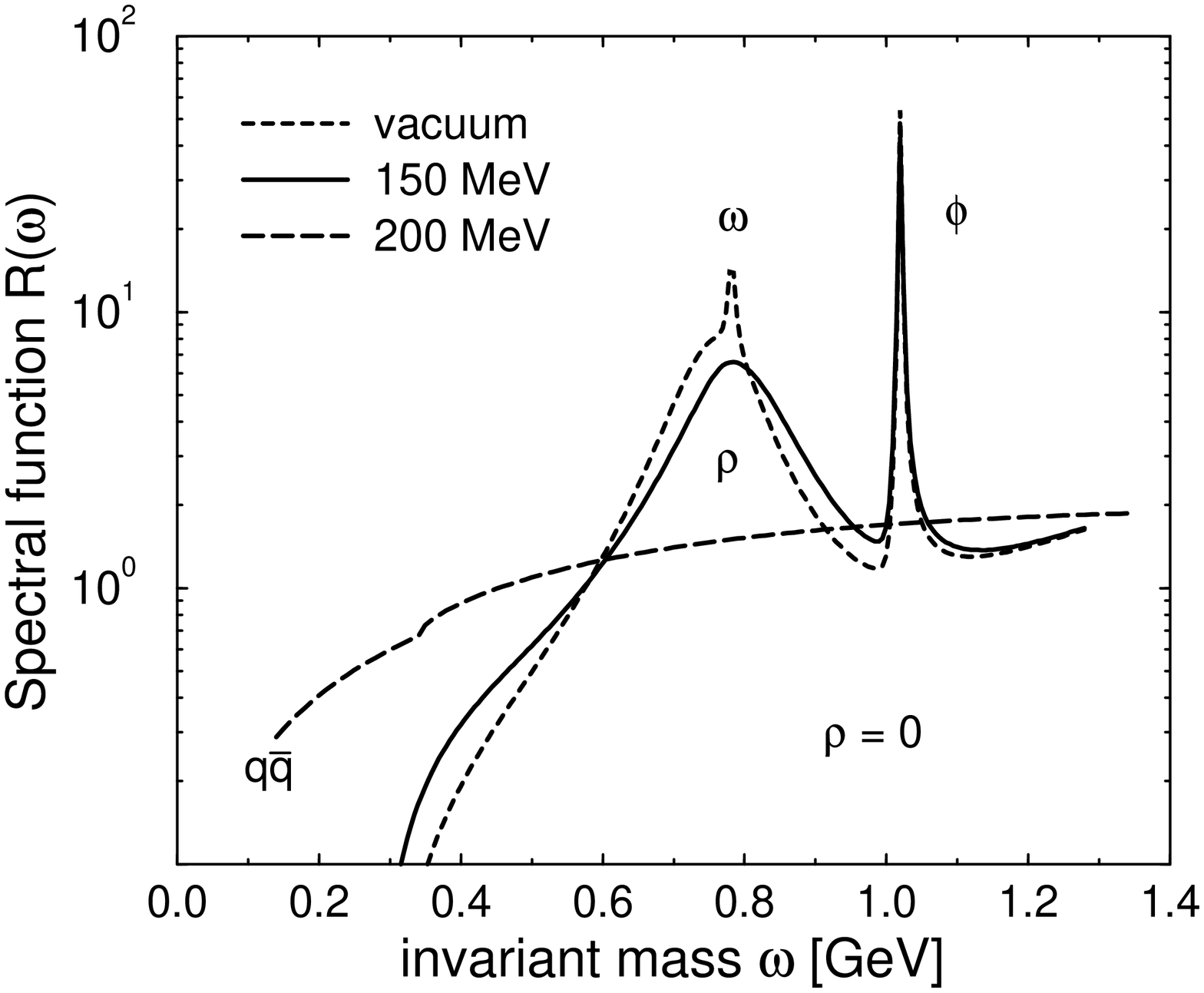,width=6.5cm, angle = -90}
\epsfig{file=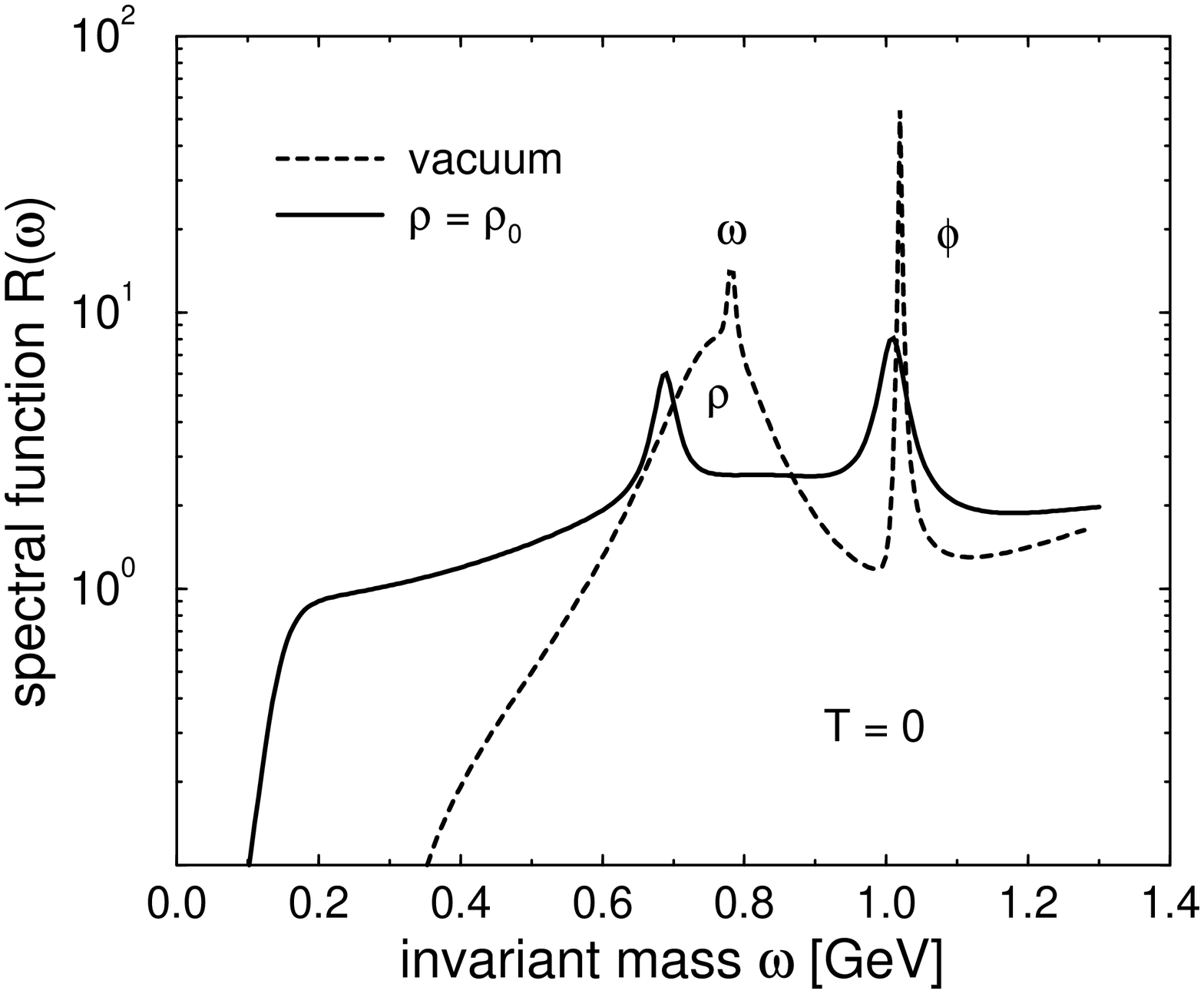,width=6.5cm, angle = -90}
\end{center}
\vspace{-1cm}
\caption{The photon spectral function $R(\omega) = -12\pi/\omega^2 \ \mbox{Im}
 \bar{\Pi}(\omega)$ at finite temperature and $\rho = 0$ (left panel), and at
$T=0$ and finite density  (right panel). The $q\bar{q}$ line in the left panel
shows the continuum spectral function in the quark-gluon phase for  massless quarks.}
\label{hadron_spectra}
\end{figure}

Finite temperature modifications of the vector meson self-energies appearing
in eq.(\ref{ImBarPi}) are calculated using thermal Feynman rules. The explicit
calculations for the $\rho$- and $\phi$-meson can be found in ref.\cite{SW00}.
At the one-loop level, the $\rho$ and $\phi$ are only marginally affected by
temperature even close to $T_C$ because of the comparably large pion and kaon
masses: $m_\pi \simeq T_C$, $m_K \simeq 3 \ T_C$. The thermal spectral
function of the $\omega$-meson has been discussed in detail in \cite{SW01}.
Here, the reaction $\omega \pi \rightarrow \pi\pi$ was found to cause a
considerable broadening of the $\omega$ spectral function, leading to
$\Gamma_\omega(T_C) \simeq 7 \ \Gamma_\omega(0)$. The resulting photon
spectral function is displayed in figure \ref{hadron_spectra} (left panel).
There is still considerable stopping of the interpenetrating nuclei at SPS
energies, with a net baryon
density $\rho_B$ in the central rapidity region. At RHIC, on the other hand,
first measurements indicate that finite baryon density effects should not play an important role. For the evaluation of density effects at SPS, we use the
results discussed in \cite{KKW2}. The photon spectral function at finite density and
zero temperature is depicted in figure \ref{hadron_spectra} (right panel).
\subsection{Fireball model}
%\begin{figure}[h]
%\begin{center}
%\epsfig{file=entropy_rhic.eps, width=7.5cm}
%\epsfig{file=entropy_sps.eps, width=7.5cm}
%\end{center}
%\vspace{-1cm}
%\caption{\label{F-EntropyRHIC} Left panel: The temperature dependence of the e%ntropy density $s$
%in the RHIC scenario as compared to the ideal quark-gluon gas and the
%noninteracting massless
%pion gas limit (dotted). The three relevant regions used in the model
%calculation are given as
%ideal hadronic resonance gas (full), interpolation (dash-dot) and quasiparticl%e
%model (dashed).  Right panel: Temperature dependence of the entropy density fo%r
%SPS conditions, including the interpolation to the fugacity corrected value
%at freeze-out. }
%\end{figure}
The fireball model is set up as follows: We assume a spatially averaged
time and density profile throughout the evolution of the fireball
at any given timeslice. Furthermore, we require the evolution of the system
after the very initial stages to be isentropic. Given the value of the total entropy $S_0$ and the
volume expansion of the fireball $V(t)$, the temperature profile
$T(t)$ can then be constructed via the EoS of the system, depending on the
relevant degrees of freedom. For temperatures above $T_C$, we use
the results of the quasiparticle model described earlier. In the region below
$T_C$, an interacting hadronic gas is an adequate description of the system.
\begin{figure}[h]
\begin{center}
\epsfig{file=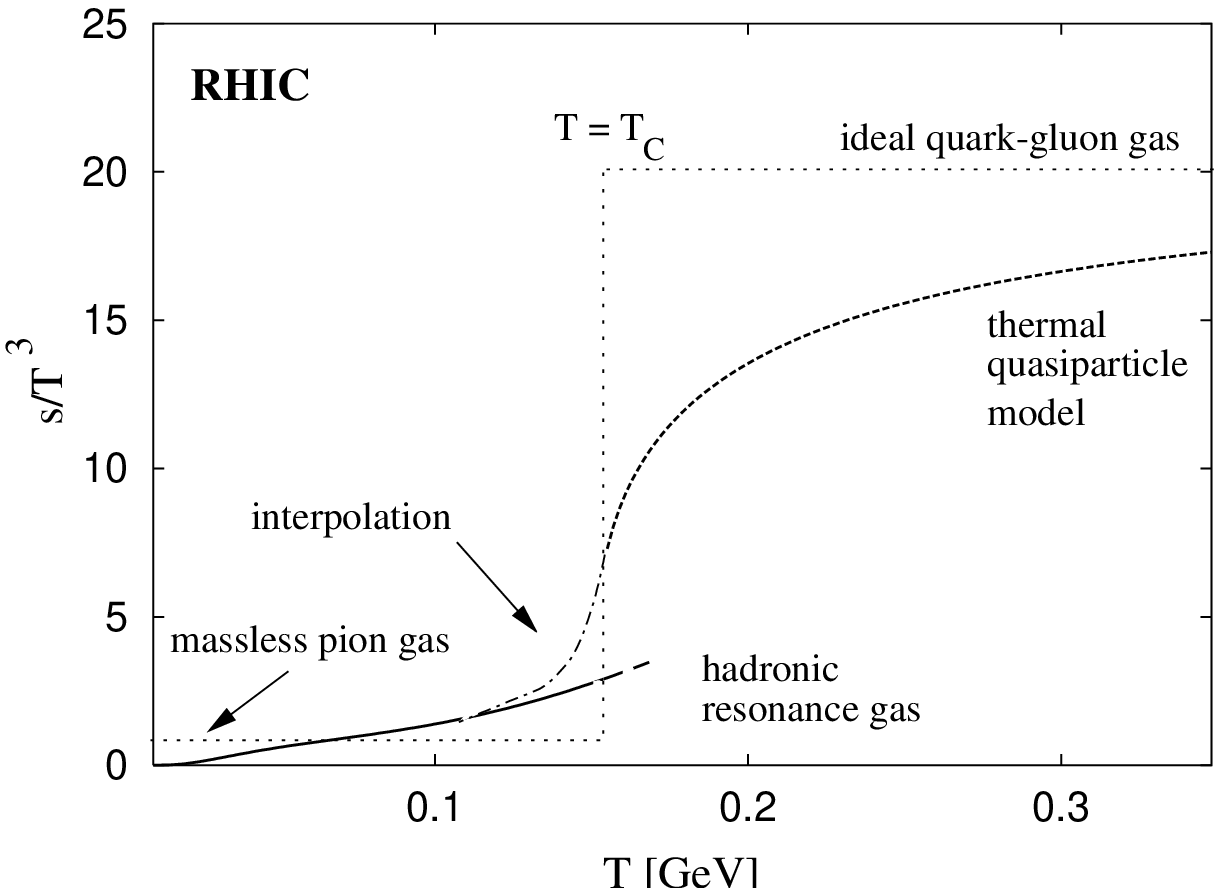, width=7.5cm}
\epsfig{file=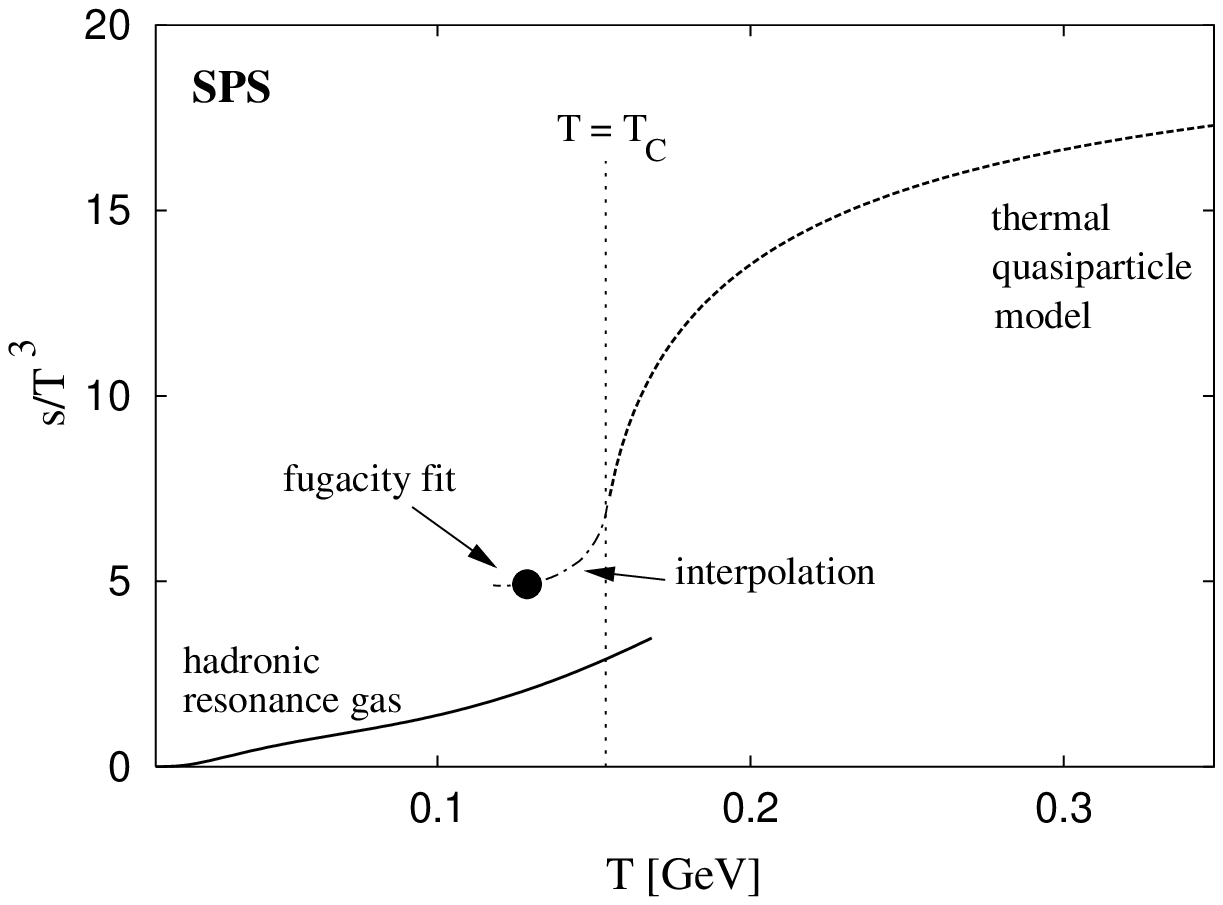, width=7.5cm}
\end{center}
\vspace{-1cm}
\caption{\label{F-EntropyRHIC} Left panel: The temperature dependence of the entropy density $s$
in the RHIC scenario as compared to the ideal quark-gluon gas and the
noninteracting massless
pion gas limit (dotted). The three relevant regions used in the model
calculation are given as
ideal hadronic resonance gas (full), interpolation (dash-dot) and quasiparticle
model (dashed).  Right panel: Temperature dependence of the entropy density for
SPS conditions, including the interpolation to the fugacity corrected value
at freeze-out. }
\end{figure}
We parametrize our insufficient knowledge
close to $T_C$  by interpolating smoothly between two regimes. This approach is supported by the general
shape of the EoS on the lattice for two light and one heavy
quark, where a smooth crossover is indicated \cite{KAR}. The entropy density in
our model is shown in figure \ref{F-EntropyRHIC}. Additionally, at SPS conditions,
finite chemical potentials for mesons and baryons have to be taken into
account. The evolution is stopped in the model as soon as
a common freeze-out temperature $T_{f} \simeq 125$ MeV is reached. Given this basic framework, we adjust the remaining parameters
of the model in such a way as to reproduce several hadronic
observables such as the rapidity distribution of the produced
hadrons, $\mathbf{p}_t$-spectra and particle ratios (for details see \cite{RSW01}).
\subsection{Results}
%\begin{figure}[tb]
%\begin{center}
%\epsfig{file=profile-sps_f.eps, width=7.5cm}
%\epsfig{file=profile-rhic_f.eps, width=7.5cm}
%\end{center}
%\caption{\label{F-TProfiles} Time evolution of the temperature for SPS and
%RHIC conditions as obtained with the fireball model described in this work.}
%\end{figure}
%
Initial temperatures are 
quite high in our approach (250 MeV for SPS conditions, 300 MeV for RHIC), and we find a
significantly prolonged lifetime ($10$ fm/c at SPS and $15$ fm/c at RHIC) of the
QGP evolution phase of the fireball as compared to the results obtained by
other groups \cite{OTHER}. This is a consequence of using the more realistic
EoS of the thermal quasiparticle model rather than the one of the ideal quark-gluon gas.
\begin{figure}[h]
\begin{center}
\epsfig{file=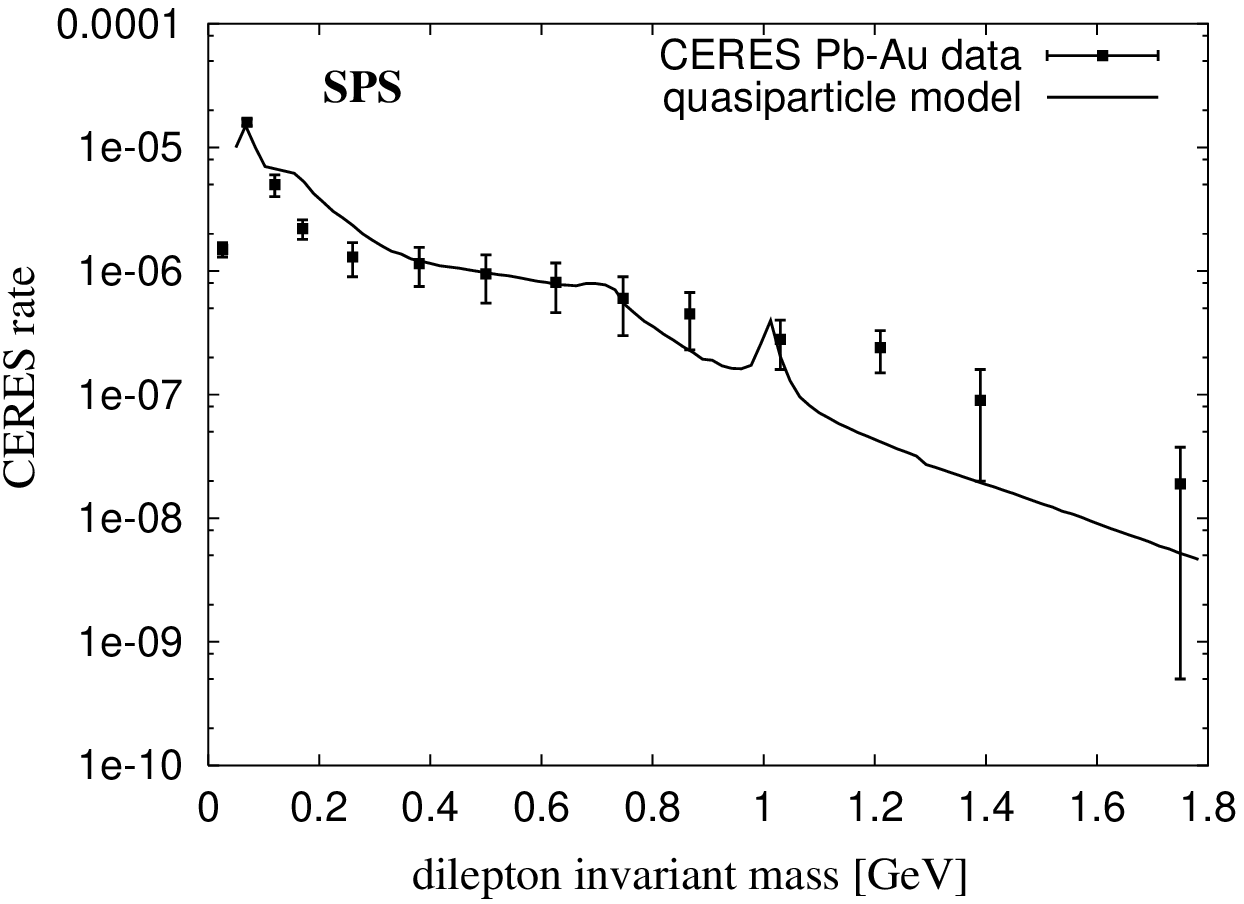, width=7.5cm}
\epsfig{file=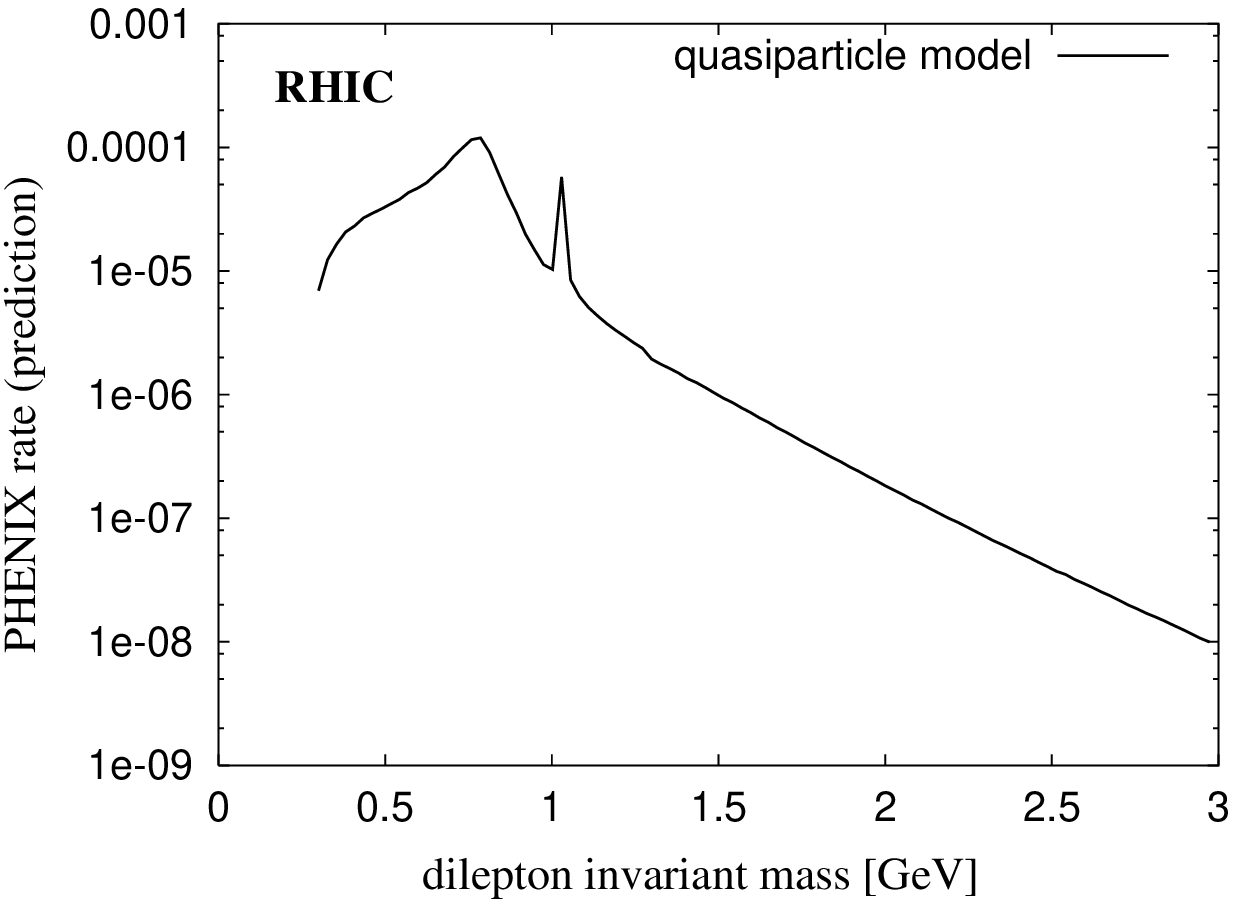, width=7.5cm}
\end{center}
\vspace{-1cm}
\caption{\label{F-Results}Dilepton invariant mass spectra for the SPS
and RHIC conditions realized in the CERES/NA45 experiment \cite{AGA} (data, upper
panel) and the PHENIX detector (prediction, lower panel).}
\end{figure}

The results for the fully integrated dilepton rates are shown in
figure \ref{F-Results}. Comparing to the CERES experiment, the model achieves a
good overall description. We are also able to describe the existing data in
separate $\mathbf{p}_t$-regions (larger or smaller than 500 MeV).

\begin{figure}[ht]
\begin{center}
\vspace*{-.8cm}
\epsfig{file=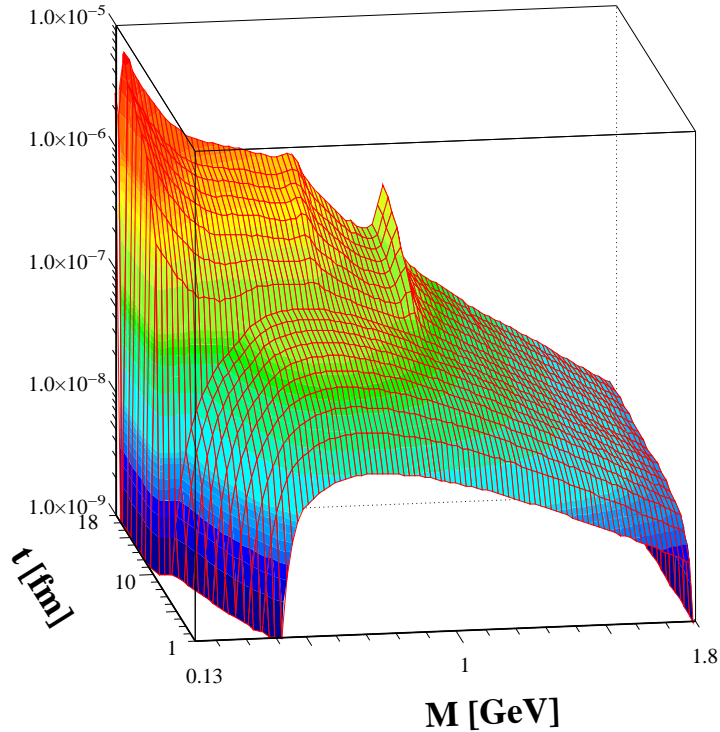, width=10cm}
\vspace{-.8cm}
\end{center}
\vspace{-1cm}
\caption{\label{F-SPSTime}Time evolution of the (integrated) dilepton yield
for the SPS scenario, as a function of dilepton invariant mass \cite{RSW01}.}
\end{figure}

In figure~\ref{F-SPSTime}, the different stages of the fireball evolution
building up the dilepton yield are resolved in small time steps.  For early times, the only contribution comes
from the $q\overline{q}$ quasiparticle anihilation processes out of the
quark-gluon phase. The movement
of the invariant mass threshold reflects the temperature dependence of the
quasiparticle mass, which gets light near the phase transition at
$t \sim 10$ fm. One observes that, in spite of the growing fireball volume,
the contributions from later timeslices to the total QGP yield become
successively less important. This behaviour is enforced by the confinement factor which
reduces the thermodynamically active degrees of freedom near the
phase boundary. For times later than 10 fm/c, the system enters the hadronic evolution phase without
going through a mixed phase. The most prominent feature is the rapid filling of
the low invariant mass region through the density-broadened $\rho$ meson.
\section{Summary}
In our journey through different regions of the QCD phase diagram, we have
first discussed the temperature and density dependence of the order parameter
of spontaneous chiral symmetry breaking, the quark condensate. The in-medium
evolution of this order parameter should manifest itself in a corresponding
dropping of the pion decay constant. This in turn may lead to observable
consequences for low-energy pion-nucleus interactions in systems with large
neutron excess. New high-precision experiments producing deeply bound states of
pionic atoms are a promising source of information to investigate these issues.

The major part of this presentation concentrated on the more extreme conditions
encountered in heavy-ion collisions at CERN and RHIC. Signals of the expansion
from a transient quark-gluon phase to the hadronic final states can in
principle be observed by detecting lepton pairs from the expanding fireball
produced in such collisions. We have emphasized the importance of using input
and constraints from lattice QCD thermodynamics in the analysis of such signals.

We thank N.~Kaiser, R.~Rapp and J.~Wambach for stimulating discussions.


\begin{thebibliography}{99}
\bibitem{GOL} M.~Golterman and S.~Peris, Phys. Rev. {\bf D 61} (2000) 034018; E.~Marco and W.~Weise, Phys. Lett. {\bf B 482} (2000) 87.

\bibitem{THOR} V.~Thorsson and A.~Wirzba, Nucl. Phys. {\bf A 589} (1995) 633;
M.~Kirchbach and A.~Wirzba, Nucl. Phys. {\bf A 604} (1996) 395; G.~Chanfray, M.~Ericson and J.~Wambach, Phys. Lett. {\bf B 388} (1996) 673.

\bibitem{WW01} W.~Weise, in: Proceedings "Nuclei and Nucleons", Darmstadt, 2000,
Nucl. Phys. {\bf A} (2001), to appear, and refs. therein.

\bibitem{KSW01} N.~Kaiser, T.~Schwarz and W.~Weise, in preparation.

\bibitem{GER} P.~Gerber and H.~Leutwyler, Nucl. Phys. {\bf B 321} (1998) 387.

\bibitem{BOY} G.~Boyd et al., Phys. Lett. {\bf B 349} (1995) 170; F.~Karsch, Nucl. Phys. (Proc. Suppl.) {\bf B 83--84} (2000) 14.

\bibitem{GIL} H.~Gilg et al., Phys. Rev. {\bf C 62} (2000) 025201; K.~Itahashi
et al., Phys. Rev. {\bf C62} (2000) 025202; T.~Yamazaki et al., Z. Physik {\bf A 335} (1996) 219.

\bibitem{WBW97} T.~Waas, R.~Brockmann and W.~Weise, Phys. Lett. {\bf B 405}
(1997) 215.

\bibitem{KW01} N.~Kaiser and W.~Weise, Phys. Lett. {\bf B 512} (2001), 283.

\bibitem{FK00} F.~Karsch, E.~Laermann, A.~Peikert, C.~Schmidt and S.~Stickan,
Nucl. Phys. (Proc. Suppl.) {\bf B 94} (2001) 411. 

\bibitem{RW01} R.A.~Schneider and W.~Weise, hep-ph/0105242, subm. to Phys. Rev. C 

%\bibitem{PKS00} A.~Peshier, B.~K\"{a}mpfer and G.~Soff, Phys.\ Rev.\  {\bf C61%} (2000) 045203.

%\bibitem{LU98} P.~Levai and U.~Heinz, Phys.\ Rev.\ {\bf C57} (1998) 1879.

%\bibitem{AG98} P.~Aurenche, F.~Gelis, R.~Kobes and H.~Zaraket, Phys. Rev. {\bf D58} (1998) 085003.

%\bibitem{AG00} P.~Aurenche, F.~Gelis and H.~Zaraket, Phys. Rev. {\bf D61} (2000) 116001.

\bibitem{KKW1} F.~Klingl, N.~Kaiser and W.~Weise, Z. Phys. {\bf A356} (1996) 193.

\bibitem{SW00} R.A.~Schneider and W.~Weise, Eur. Phys. J. {\bf A9} (2000) 357.

\bibitem{SW01} R.A.~Schneider and W.~Weise, Phys. Lett. {\bf B} (2001), in print.

\bibitem{KKW2} F.~Klingl, N.~Kaiser and W.~Weise, Nucl. Phys. {\bf A624} (1997) 527.

\bibitem{KAR} F.~Karsch, plenary talk, Proceedings QM2001, to be published.

\bibitem{RSW01} T.~Renk, R.A.~Schneider and W.~Weise, preprint, submitted to
Phys. Rev. {\bf C}.

\bibitem{OTHER} R.~Rapp and J.~Wambach, Eur.\ Phys.\ J.\ {\bf A 6} (1999) 415; J.~Sollfrank, P.~Huovinen, M.~Kataja, P.~V.~Ruuskanen, M.~Prakash and R.~Venugopalan,
Phys.\ Rev.\ {\bf C 55} (1997) 392; C.~M.~Hung and E.~Shuryak, Phys.\ Rev.\
{\bf C 57} (1998) 1891.

\bibitem{AGA} G.~Agakichiev et al., Phys. Lett. {\bf B 422} (1998) 405 and
refs. therein.

\end{thebibliography}
\end{document}